\pdfoutput=1
\documentclass[11pt,usletter]{article}
\usepackage{jheppub}
\usepackage[section]{placeins}
\usepackage{graphicx}
\usepackage{hyperref}
\usepackage{color}

\def\ba{\begin{align}}
\def\ea{\end{align}}
\def\be{\begin{equation}}
\def\ee{\end{equation}}
\def\bea{\begin{eqnarray}}
\def\eea{\end{eqnarray}}

\begin{document}

\title{Artificial Neural Network in Cosmic Landscape}

\author[a]{Junyu Liu}
 \affiliation[a]{Walter Burke
  Institute for Theoretical Physics, California Institute of
  Technology, Pasadena, California 91125, USA}
\emailAdd{jliu2@caltech.edu}

\abstract{In this paper we propose that artificial neural network, the basis of machine learning, is useful to generate the inflationary landscape from a cosmological point of view. Traditional numerical simulations of a global cosmic landscape typically need an exponential complexity when the number of fields is large. However, a basic application of artificial neural network could solve the problem based on the universal approximation theorem of the multilayer perceptron. A toy model in inflation with multiple light fields is investigated numerically as an example of such an application.}


\maketitle
\flushbottom

\section{Introduction}
Inflation is one of the most promising cosmological paradigms to describe the dynamics of spacetime and quantum fields in the very early universe \cite{Guth:1980zm,Linde:1981mu,Albrecht:1982wi}, although many details of inflationary dynamics are still mysterious and are not confirmed by theories and observations \cite{Martin:2013tda}. For instance, we are still curious that, if inflation is governed by scalar fields, what is the corresponding inflationary potential? Indicated by string theory\footnote{The study of the string theory vacua \cite{Candelas:1985en,Kachru:2003aw,Susskind:2003kw,Vafa:2005ui,ArkaniHamed:2006dz,Douglas:2004kp,Douglas:2004zg,Douglas:2006za} shows that there might exist a cosmic landscape with large degrees of freedom during the very early universe.}, there should be a complicated vacuum structure and a tower of light scalar fields is predicted and might drive the dynamics of cosmic inflation. Theoretically, we might treat the inflationary potential
\begin{align}
V(\varphi_j)=V(\varphi_1,\varphi_2,\varphi_3,\cdots,\varphi_n)
\end{align}
to be highly random \cite{Berera:1996nv,Huang:2008jr,Tye:2009ff,Battefeld:2011yj,Dias:2012nf,Green:2014xqa} on the field space.
\\
\\
A directly way of studying physical properties of inflation governed by such a random potential is computer simulation (although some specific problems could also be answered with mathematics, for instance, studying statistics using Kac-Rice formula, see \cite{Bardeen:1985tr,Easther:2016ire}). In order to realize this idea in a real computer and get some predictions, we typically need to use Monte-Carlo method to generate a random function up to some measures, and try to solve the dynamical equation of fields in inflation. The methods to investigate such a problem in the existing research can be classified in two types: global and local methods (see also in \cite{Liu:2015dda}). The advantages and disadvantages are given as following.
\\
\\
The local type of methods is trying to generate the random function $V(\varphi_j)$ around some specific inflationary trajectories $\varphi_j(t)$, namely, solving the differential equation at the same time when generating random functions (where their work typically established on random matrix theory or statistical physics, see \cite{Aazami:2005jf} for early contribution on using random matrix theory in cosmic landscape, and also \cite{Marsh:2013qca,Battefeld:2014qoa} for locally constructing inflationary trajectories). In each time step, one can generate a random potential in a small neighbourhood for the corresponding point in the field space, and then use that potential to evolve the time towards the next step by treating the differential equations as difference equations, and then generate a new, consistent random potential at the second step and so on. Redo this algorithm for several times, one can obtain an inflationary trajectory. This type of methods is often very fast to generate a high-dimensional landscape with polynomial computational complexity, and capture most of physics related to inflationary trajectories. However, lacking of a global structure of the landscape may lead to an inconsistency for self-crossing trajectories without special treatment \cite{Bachlechner:2014rqa}, which might make it harder to study some specific phenomena in inflation, such as periodic, self-crossing trajectories and bifurcation.
\\
\\
The global type is trying to directly generate the whole random function $V$ in a compact or non-compact region of field space. The related technologies include Fourier transformation \cite{Tegmark:2004qd,Frazer:2011br} and interpolation \cite{Duplessis:2012nb}. This method has no inconsistency between different trajectories because the landscape is global, and it is safe to study special phenomena which need the consistency in the space of trajectories, like self-crossing. However, it is hard to discuss a high-dimensional landscape (which might be more realistic according to string theory vacua) because the computational complexity is exponential if we don't use further assumptions \cite{Liu:2015dda}.
\\
\\
In this paper we argue that, artificial neural network, the basic construction of machine learning in the computer science \cite{book1,book2,book3,ANN}, is naturally suitable for globally generating a landscape. Based on the celebrated universal approximation theorem \cite{approx,approx2}, artificial neural network is a universal function simulator. Thus introducing the randomness in the parameters of artificial neural network will generate all possible random functions in the corresponding range. For a given construction of artificial neural network, the computational complexity is polynomial with field degrees of freedom, up to some controlled parameters labeling the randomness (or capacity) of the output. Then we will do some numerical investigations on the landscape generation, and try to evolve the inflationary trajectories based on the artificial-neural-network-generated landscape in an inflationary model.
\\
\\
This paper will be organized as following. In Section \ref{RF}, after a discussion on computational complexity of traditional algorithms, we will discuss artificial neural network and its application on the landscape. In Section \ref{Nu}, we will present some numerical tests of landscape constructed by artificial neural network, and possible applications for evolving inflationary trajectories. In Section \ref{conc}, we will arrive at a conclusion and some outlooks on the future directions.
\section{Random function and artificial neural network}\label{RF}
\begin{figure}[htbp]
  \centering
  \includegraphics[width=0.32\textwidth]{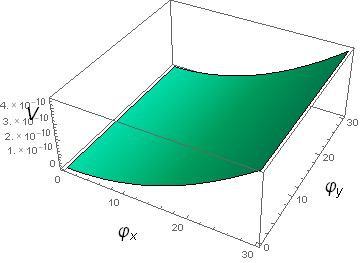}
  \includegraphics[width=0.32\textwidth]{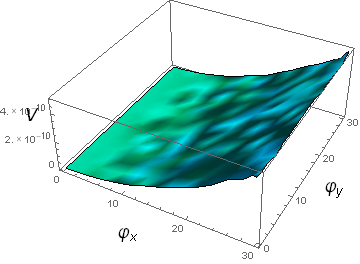}
  \includegraphics[width=0.32\textwidth]{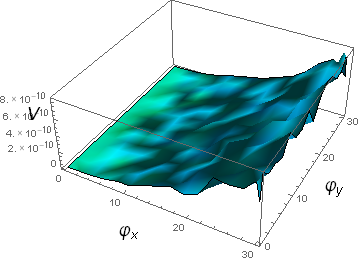}
  \caption{\label{fig:interpolation} Potential surface with null, small and large randomness. We choose $\Delta_x\varphi=0.05$, $m=10^{-6}$, $\varphi_{\max}=30$, and $\mathcal{A}=0,0.1,1$ respectively for the left, middle and right figure. The distribution is uniform for random numbers in the interpolation function, and the interpolation is for $C^1$ functions.}
\end{figure}

In order to globally construct a random function, the most direct method might be interpolation. The algorithm can be simply described as following (for simplicity we assume the dimension $n=1$). Consider smooth functions on the interval $I=[a,b]$. First, we split the domain $[a,b]$ into $N$ small intervals. For each interval $I_i$, we generate a random number $f_i$ living in some specific distributions (Gaussian or uniform distribution). Finally, we do the interpolation of series $(a_i,f_i)$, where $a_i$ is a characteristic point of $I_i$ (eg, we can choose $a_i$ as the middle point of $I_i$). After interpolation, we obtain a smooth curve, which can be regarded as a one-dimensional random function construction living in the function space $C^m[a,b]$ (where $m$ depends on the interpolation algorithm we choose).
\\
\\
This method can be easily generalized to high-dimensional case, where the interpolation and sub-intervals should be high-dimensional. For instance, we can interpolate a smooth curve as a two-dimensional random function where two-dimensional domain is split to many small boxes. Moreover, we can interpolation on one fixed surface with some random bumps. In the following example, we choose the quadratic potential surface on the domain $[0,\varphi_{\max}]^2\subset \mathbb{R}^2$
\begin{align}\label{basisf}
V({{\varphi }_{x}},{{\varphi }_{y}}) =\frac{1}{2}m^2\varphi_x^2
\end{align}
and after random perturbations, we have the interpolation
\begin{align}
V_{ij}=\frac{1}{2}(1+\mathcal{A}\operatorname{Rand}_{ij})m^2(i\Delta_x\varphi)^2
\end{align}
where $\Delta_x\varphi=\frac{\varphi_{\max}}{N}$ is the distance for each adjacent pair of two-dimensional intervals, $\operatorname{Rand}_{ij}$ are random numbers in $[0,1]$ chosen for each $(i,j)$ and belong to one fixed distribution, and $\mathcal{A}$ is the amplitude of randomness. This function is made to be asymmetric in $i$ and $j$ (corresponds to $x$ and $y$), because the mean part of this random function is driven by the one-dimensional variable $x$ (like eq.\ref{basisf}), while some small two-dimensional random perturbations are around such a function. (In cosmology, one can imagine that it is a two-dimensional inflationary model with one inflaton and another light field which will not drive inflation. The interaction between two fields is due to a random landscape). Figures with different randomness are shown as Figure \ref{fig:interpolation}.
\\
\\
However, this method is practically useless for high-dimensional case. Just see the number of random variables related on the dimension $n$. If we choose $N$ intervals on each direction, we will obtain $N^n$ high-dimensional intervals. As a result, the number of random variables are $N^n$. And we must do high-dimensional interpolation on these $N^n$ random numbers. So the complexity is exponentially blowing up with the dimension $n$.
\\
\\
The same situation happens on other global methods. Taking classical Fourier series for example, the one-dimensional Fourier expansion is
\begin{align}
f(x)=\sum\limits_{\ell=1}^{N}{{{a}_{\ell}}{{e}^{i\ell x}}}
\end{align}
So we have $N$ Fourier coefficients (after truncation from infinite sum for a practical computation in the computer). The two dimensional Fourier expansion is
\begin{align}
f({{x}_{1}},{{x}_{2}})=\sum\limits_{{{\ell}_{1}}{{\ell}_{2}}}^N{a_{{{\ell}_{1}}}^{(1)}a_{{{\ell}_{2}}}^{(2)}{{e}^{i{{\ell}_{1}}{{x}_{1}}}}{{e}^{i{{\ell}_{2}}{{x}_{2}}}}}
\end{align}
So we have $N^2$ Fourier coefficients after truncation. Generally, the $n$ dimensional Fourier expansion is
\begin{align}
f({{x}_{1}},{{x}_{2}},\ldots ,{{x}_{n}})=\sum\limits_{{{\ell}_{1}}{{\ell}_{2}}\ldots {{\ell}_{n}}}^N{\prod\limits_{j=1}^{n}{a_{{{\ell}_{j}}}^{(j)}{{e}^{i{{\ell}_{j}}{{x}_{j}}}}}}
\end{align}
So we will have $N^n$ Fourier coefficients. When we want to use classical Fourier series to simulate the high-dimensional random function, we must set all the coefficients to be random numbers and the computational complexity will be $\mathcal{O}(e^n)$. In the inflationary landscape, the number of light fields could be larger than $\mathcal{O}(10)$-$\mathcal{O}(100)$. Thus, we need a new way to construct high-dimensional random functions.
\\
\\
Recently there is a rising interest in physics to apply computer science technologies into physical problems (for instance, see \cite{Harlow:2013tf,Denef:2006ad,Denef:2017cxt,Bao:2017thx}). As one of the most promising area in computer science, machine learning (or artificial intelligence) \cite{book1,book2,book3} is widely used to understand difference areas of physics, including condensed matter phases \cite{apply1}, high energy experiment \cite{Aad:2014yva}, energy landscape \cite{machine1,machine2,Ballard:2017agc}, particle phenomenology \cite{Cohen:2017exh}, tensor networks \cite{tensor} and cosmic non-Gaussianities \cite{Novaes:2014ska}. In recent research string theorists also find that machine learning algorithm is efficient to study manifold data in the string landscape \cite{He:2017aed,Krefl:2017yox,Ruehle:2017mzq,Carifio:2017bov}, which may give us the motivation to think about the learning algorithm landscape from cosmological point of view.
\\
\\
Artificial neural network is a fundamental construction in modern machine learning. Inspired by the research of brain and neural biology, artificial neural network is proposed as a mathematical model to simulate the activity of calculation and identification of human brains, as highly non-linear, complicated biological computers. Mathematically speaking, artificial neural network is given by some specific linear combinations of activation function. A mathematical proof is given to claim that all possible continuous functions on a compact space could be simulated by artificial neural networks \cite{approx,approx2}. By learning algorithms, one can tune the input parameters in the network to realize some computational objects efficiently, like pattern recognition.
\\
\\
The unit in artificial neural network is called neuron. A neuron is mathematically defined as a function mapping a vector to a scalar. It is combined by the following objects,
\begin{itemize}
\item \emph{Synapse}. Synapses are typically connected by some weights. To be concrete, an input signal for the synapse and connected to a neuron, which is called $x_j$, is multiplied by a corresponding weight $w_j$.
\item \emph{Adder}. Adder is the addition operation of all input signals.
\item \emph{Activation function}. An activation function $\psi$ is used to model the output of the neuron from external influence. Typical types of activation functions include sigmoid function or Heaviside function, which are widely used in machine learning. The function should often be assumed as nonconstant, bounded, and monotonically-increasing continuous, as the requirement of the universal approximation theorem. However, some periodic functions, like sine function, are also used to be the approximator. (For instance, see \cite{thesis}.)
\item \emph{Bias}. The bias $b$ is defined as a constant shift for input signals.
\end{itemize}
To be specific, for an $n$-dimensional signal vector $\{x_j\}_{j=1}^n$, the neural output $y$ is defined as
\begin{align}
y=\psi \left( \sum\limits_{j=1}^{n}{{{w}_{j}}{{x}_{j}}}+b \right)
\end{align}
A simple example of artificial neural network is the \emph{multilayer perceptron}. In order to realize some computational objects, we need to use a well-defined, organized, layered neural network of multiple neurons, which should have at least two layers: the output and input layers of neurons. For some complicated constructions, their might be also some layers in the middle, which are called hidden layers. In the simplest construction, signals are moving in only one direction, namely the connections between units will not form cycles. This is called feed-forward type, or multilayer feed-forward architecture. If some of the activation functions are non-linear, it is called the multilayer perceptron.
\\
\\
More hidden layers will significantly increase the quality of statistics as the feedback of stimulation for an artificial neural network. In a real network which is designed for complicated computational tasks, the number of outputs, inputs and hidden layers is large. In our practical motivation, we are only interested in the random function, thus we will consider a multilayer perceptron with one output, and we will only consider one hidden layer for simplicity. Namely, if we have a multilayer perceptron neural network with one hidden layer that consists of $h$ hidden units, and the network has $n$ inputs, then we can formalise the output of the network as a function of the inputs by
\begin{align}\label{in}
{{A}_{h}}({{x}_{1}},{{x}_{2}},\ldots ,{{x}_{n}})=\sum\limits_{j=1}^{h}{{{w}_{j}}\psi }\left( \sum\limits_{i=1}^{n}{{{a}_{ij}}{{x}_{i}}}+{{b}_{j}} \right)
\end{align}
where $a_{ij}$ is the weight of the synapse that goes form the input $x_i$ to the $j$th hidden neuron, $b_j$ is the bias of the $j$th hidden neuron, $\psi$ is the activation function, and $w_i$ is the weights of the synapse that goes to the output neuron to form the $j$th neuron.
\\
\\
For this configuration, we have the simplest \emph{universal approximation theorem}. The universal approximation theorem shows that the standard multilayer feed-forward networks with arbitrary activation functions, are universal approximators for all possible function in the corresponding variable dimension and the compact domain. This property is not because of the form of activation functions, but the structure of multilayer perceptron itself \cite{approx,approx2}. The theorem claims that,
\\
\\
\emph{
Let $\psi$ be the activation function. Let $X^n\subset\mathbb{R}^n$ and $X^n$ be compact. Then for $\forall k=1,2,\ldots,\infty$, for $\forall f \in C^k[X^n]$, for $\forall \epsilon>0$: $\exists h\in\mathbb{N}$, $a_{ij}\in\mathbb{R}$, $b_j\in\mathbb{R}$, $w_j\in\mathbb{R}$, $i\in{\{1,2,\ldots,n\}}$, $j\in{\{1,2,\ldots,h\}}$, s.t.:
\begin{align}
\text{for }\forall \vec{x}=({{x}_{1}},{{x}_{2}}\ldots ,{{x}_{n}})\in {{X}^{n}}:\left| f(\vec{x})-{{A}_{h}}(\vec{x}) \right|<\epsilon
\end{align}}
\\
This theorem could be regarded as the theoretical foundation of artificial neural networks, and could be easily generated to more complicated cases. According to this theorem, properly choices of parameters will lead to an efficient approximation to all possible high-dimensional functions. Thus, including the randomness in the parameters will give a nice and efficient construction of random functions.
\\
\\
About specific choice of activation functions, the original theorem needs the activation function to have some specific properties, to be concrete, bounded and monotonically-increasing. However, some other types of functions could also be used in the practical construction of neural networks. In the following application, we will try sigmoid function
\begin{align}
\psi (x)=\frac{1}{1+{{e}^{-x}}}
\end{align}
and sine function
\begin{align}
\psi (x)=\sin(x)
\end{align}
A related idea also appear in the deep learning and statistical physics, which is searching for patterns in some statistical models energy landscape (for instance, see \cite{machine1,machine2,Ballard:2017agc}). There problem is related to the solution of a minimal (local or global) in a high dimensional surface (where is called \emph{landscape}), or specifically finding a minimal energy in some models from statistical physics, and the high dimension comes from the large $N$ setup in the corresponding model. The energy landscape, especially in higher dimensions, are very hard to explore globally. However, the task could be related to machine learning algorithms and might be solved efficiently. These algorithms might benefit the future research of string landscape for solving some statistical patterns.

\section{Construction and application}\label{Nu}
\subsection{Random function generation}
\begin{figure}[htbp]
  \centering
  \includegraphics[width=0.4\textwidth]{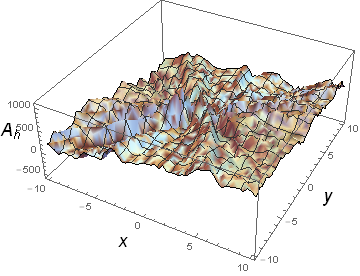}
  \includegraphics[width=0.4\textwidth]{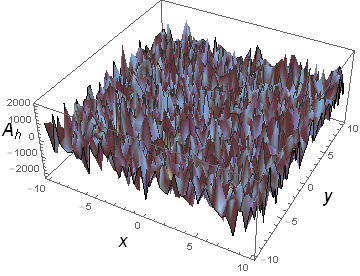}
  \caption{\label{fig:randslice} Slices of two 100-dimensional random functions. We generate 100-dimensional random functions with parameters $\alpha=50$, $\beta=50$, $\omega=50$, $h=1000$ for sigmoid (left) and sine (right) activation functions respectively. We plot the function $A_h(0,0,\ldots,x,y)$ (The first 98 components of the variable vector are all zero and the last two components are chosen to be $x$ and $y$.)}
\end{figure}

From the universal approximation theorem mentioned in the last subsection, one con approximate any arbitrary continuous or $k$th order continuous function by choosing different parameters $h,a_{ij},b_j,w_j$. As a result, if we choose these parameters randomly, we can construct a smooth random function living in some functional distribution. This is the basic idea of our random generation method.
\\
\\
Some comments are given as following.
\begin{itemize}
\item  The number of neurons in the hidden layer $h$, should be fixed in a practical generation. In fact, $h$, and also the number of hidden layers for more complicated networks, can be roughly regarded as the capability of random statistics it could generate. Larger $h$ could always in principle reproduce the result for lower $h$. $h$ could be regarded as an adjustable variable for users according to the computational ability of their system.
\item  The range of parameters $a_{ij},b_j,w_j$ should be bounded, and the output signals will live in a given range. For instance, one can assume
\begin{align}
a_{ij}\in[-\alpha, \alpha] \ \ b_j\in[-\beta,\beta] \ \ w_j\in[-\omega,\omega]
\end{align}
for uniform distributions.
\item The generation is fast. We can do a simple analysis on the time scale it will take. If we assume that when we compute $\varphi(x)$ once for some $x$, we need the time $t_1$; when we compute the addition or multiplication once, we need the time $t_2$; when we generate a random number, we need $t_3$, then for a total construction process of one random function, we need time
\begin{align}
hd(2{{t}_{2}}+{{t}_{3}})+h({{t}_{1}}+2{{t}_{2}}+2{{t}_{3}})-{{t}_{2}}=\mathcal{O}(hd)
\end{align}
for one hidden layer. For finite number of hidden layers, one sample random function could be generated in a polynomial time.
\end{itemize}
In Figure \ref{fig:randslice} we show slices of a 100-dimensional random function generated by multilayer perceptron with one hidden layer.
\subsection{Properties}
In the previous sections, we argue that artificial neural network provides a new effective approach to simulate a landscape. Thus, it is valuable to obtain basic properties of random functions generated by neural networks. Some analytic and numerical results are summarized in the following discussions.
\subsubsection{Volume and symmetry in the field space}
Because of finiteness of digital numbers in the computer, it is typically computational expensive to explore the field range in full $\mathbb{R}^n$, or an extremely large compact subspace of it (and this is also not allowed by the universal approximation theorem). Considering practical applications, in the context of inflationary cosmology, some trans-Planckian arguments \cite{Martin:2000xs} or the Lyth bound \cite{Lyth:1996im}, may constrain the range of primordial fields and their perturbations. Thus, it should be crucial to constrain the valid volume where random functions could appear and fluctuate in the field configuration space.
\\
\\
In the traditional method, one can try to cut off the volume by brute force. For instance, in the low dimensional interpolation approach, one can choose the range of field space for the corresponding random sampling. In the Fourier method, one can also directly choose the field range. The range of field that has been cut off is typically the range where the random functions could appear. For Taylor expansion, there are some clear asymptotic dependence at infinity of the field space, thus it should be necessary to cut out the boundary asymptotic range to remain enough randomness.
\\
\\
In the context of artificial neural networks, the volume in the field space strongly depends on the activation function we choose. A simple observation is, for the sine activation function, the field space is always full, because it is periodic, fluctuating from $-1$ to $1$. The norm of fluctuation is set by the range of random number $w_j$ only (namely, $\mathcal{O}(1)\times w$), in the simplest one layer of hidden neurons. However, the sigmoid function has some different properties. It will quickly remain constant if the input is away from the origin. As a rough estimation, go to the negative direction from the origin, the amplitude decays to $1/e$ when $x=-\log(-1 + e)=-0.54=\mathcal{O}(1)$. Thus, the function is activated around a bounded regime by an $\mathcal{O}(1)$ number. As an estimation, consider the neural network construction, the activated regime of one layer of hidden neurons is
\begin{align}
&\psi ({{a}_{ij}}{{x}_{i}}+{{b}_{j}})\text{ is activated}\Rightarrow {{a}_{ij}}{{x}_{i}}+{{b}_{j}}\sim \mathcal{O}(1)\nonumber\\
&\Rightarrow {{x}_{i}}\sim \frac{\beta }{\alpha n}\Rightarrow \text{ Volume}={{(\frac{\beta }{\alpha n})}^{n}}
\end{align}
For sufficiently large number of hidden neurons.
\\
\\
Sometimes people may consider field space with some symmetric structure, for example, rotational invariance (for cosmological applications, sometimes we consider scalar fields and their perturbations may be separated as inflaton and isocurvatons with radial and angular directions in the two dimensional or higher dimensional field space \cite{Chen:2009zp}). In this case, the neural network construction should be slightly modified. A possible approach would be modify the matrix $a_{ij}$ in the network eq.\ref{in} to be rotational invariant, namely, sampling with a proper measure which is invariant under the rotation group $\text{SO}(n)$. Namely for measure $\mu(\mathbf{a}')$ on a $h\times n$ matrix, where $\mathbf{a}'=(a_{ij})^T$ in eq.\ref{in}, we have
\begin{align}
\mu(\mathbf{a}'\cdot\mathbf{g})=\mu(\mathbf{a}')\text{   where   }\mathbf{g}\in \text{SO}(n)
\end{align}
The $\text{SO}(n)$ invariant measure construction on a $h\times n$ matrix is a well-defined problem in harmonic analysis and measure theory. Here we only show the solution of the $h=n$ case (it is definitely possible for a practical simulation, where $h$ and $n$ are both large numbers), where in this case we can directly use an \emph{uniform} measure on $\text{SO}(n)$ \cite{mathbook}. The only measure which is left and right invariant, and should also be normalized, is the celebrated Haar measure \cite{Haar}, which is widely used in applied mathematics and theoretical physics, and more proper for our current motivation. The Haar measure construction for $\text{SO}(n)$ is defined as follows. For an $n$ dimensional unit sphere $S^n$, the hypersurface coordinate, denoted by $\mathbf{\Sigma}_n({{\theta }_{1}},{{\theta }_{2}}\ldots {{\theta }_{n}})$, is defined as
\begin{align}
{{\Sigma }_{n,i}}({{\theta }_{1}},{{\theta }_{2}}\ldots {{\theta }_{n}})=\left\{ \begin{matrix}
   \sin {{\theta }_{1}}\sin {{\theta }_{2}}\ldots \sin {{\theta }_{n}} & i=1  \\
   \cos {{\theta }_{i-1}}\sin {{\theta }_{i}}\ldots \sin {{\theta }_{n}} & 2\le i\le n  \\
\end{matrix} \right.
\end{align}
where $\theta_i$ are angular coordinates on $S^n$, compactified by
\begin{align}
0 \le {{\theta }_{1}}\le 2\pi  ~~~~~~~~~0\le {{\theta }_{i}}\le \pi ~~~~~ (2\le i\le n)
\end{align}
Based on the construction of $\mathbf{\Sigma}_n$, we define the following matrix
\begin{align}
&{{M}_{n+1}}({{\theta }_{1}},{{\theta }_{2}}\ldots {{\theta }_{n}})=\nonumber\\
&\left( {{\mathbf{U}}_{n,1}}({{\theta }_{1}},{{\theta }_{2}}\ldots {{\theta }_{n}}),{{\mathbf{U}}_{n,2}}({{\theta }_{1}},{{\theta }_{2}}\ldots {{\theta }_{n}}),\ldots ,{{\mathbf{U}}_{n,n}}({{\theta }_{1}},{{\theta }_{2}}\ldots {{\theta }_{n}}),{{\mathbf{\Sigma }}_{n}}({{\theta }_{1}},{{\theta }_{2}}\ldots {{\theta }_{n}}) \right)
\end{align}
where
\begin{align}
{{\mathbf{U}}_{n,i}}({{\theta }_{1}},{{\theta }_{2}}\ldots {{\theta }_{n}})=\frac{1}{\sin {{\theta }_{i}}\ldots \sin {{\theta }_{n}}}\frac{\partial {{\mathbf{\Sigma }}_{n}}({{\theta }_{1}},{{\theta }_{2}}\ldots {{\theta }_{n}})}{\partial {{\theta }_{i}}}
\end{align}
The group manifold $n(n-1)/2$ different coordinates, where we denote by ${{\left( {{\phi }_{i,j}} \right)}_{1\le i\le j\le n-1}}$ and it is compactified by
\begin{align}
0 \le {{\phi }_{1,j}}\le 2\pi ~~~~~ (1\le j\le n-1)  ~~~~~~~~~0\le {{\phi }_{i,j}}\le \pi ~~~~~ (2\le i\le j\le n-1)
\end{align}
The Haar parametrization of the matrix from $\text{SO}(n)$ is defined by recursion. Start from
\begin{align}
{{\mathbf{a}}_{2}}({{\phi }_{11}})=\left( \begin{matrix}
   \cos {{\phi }_{11}} & \sin {{\phi }_{11}}  \\
   -\sin {{\phi }_{11}} & \cos {{\phi }_{11}}  \\
\end{matrix} \right)
\end{align}
for ${\mathbf{a}}_{2} \in \text{SO}(2)$, we define ${\mathbf{a}}_{n} \in \text{SO}(n)$ by recursive matrix product
\begin{align}
{{\mathbf{a}}_{n}}\left( {{\left( {{\phi }_{i,j}} \right)}_{1\le i\le j\le n-1}} \right)={{M}_{n}}\left( {{\left( {{\phi }_{1,j}} \right)}_{1\le j\le n-1}} \right)\mathbf{a}_{n-1}^{\text{ext}}\left( {{\left( {{\phi }_{i,j}} \right)}_{1\le i\le j\le n-2}} \right)
\end{align}
where the operation $\text{ext}$ means that
\begin{align}
{{({{v}_{1}},{{v}_{2}},\ldots ,{{v}_{n-1}})}^{\text{ext}}}\equiv ({{v}_{1}},{{v}_{2}},\ldots ,{{v}_{n-1}},0)
\end{align}
mapping vectors in $\mathbb{R}^{n-1}$ to a hyperplane subspace in $\mathbb{R}^{n}$. Sampling from the uniform distribution of angles, we obtain the Haar random matrix samples. Obviously the computational complexity for this generation algorithm is polynomial in $n$. Using this method, we can obtain $a_{ij}$ from Haar measure, multiplying with an overall constant $\alpha$ to control the random range, we obtain an rotational symmetry $\text{SO}(n)$ in the field space by construction.
\\
\\
Some other methods may also achieve this goal, for instance, decomposing the field contents to radial and angular pieces. And those discussions could also be extended to some other symmetry groups. We leave the detailed discussions to future research.
\subsubsection{Taylor statistics}
The number of degree of freedom increases polynomially in the neural networks, while exponentially in the Taylor or Fourier expansion. Thus, the random Taylor or Fourier series coefficients might be correlated from a neural network random function, especially when the number of hidden neurons is small enough. This fact will not effect the universal approximation statement, because it is natural to imagine that one coefficient set from corresponding complete basis could be non-trivially transforms to another coefficient set with a different complete basis, while the transformed coefficients have some induced correlations. However, testing this property from a neural network random function is also helpful especially when we are dealing with some specific real problems where independence of some given coefficients could play a crucial rule, and then we have to modify the structure and the activation of the neural network, or change the distribution of random elements in a network construction.
\\
\\
This problem is hard to answer universally from analytic and numerical point of view. From analytic side, it is hard to control the network especially for an arbitrary activation function setup, while from numerically point of view, it is significantly hard to test the network with higher dimensions through Taylor or Fourier decomposition (because of exponential growth \footnote{Typically, the integral basis (Fourier) decomposition is even harder than derivative basis (Taylor) decomposition, because when performing an integral over a random function, the strong fluctuations make the integral hard to converge.}).
\\
\\
However, in this paper we will try to deal with this problem in the simplest setup, namely, considering the single layer network with $\mathcal{O}(100)$-$\mathcal{O}(1000)$ hidden layers for a two dimensional random function, and calculating the Taylor series. We consider $1000$ random realizations to get some statistical predictions. Given a random function, if we expand through Taylor series, an important feature is that we often obtain a non-perturbative series, namely, the coefficients from higher orders are crucially larger than the lower cases. This is because the function is highly random such that the local Taylor expansion is only valid in a small regime, which is divergent outside this regime. The non-perturbative coefficients might be highly random and strongly fluctuating over a very large regime, while the lower order coefficients are relatively stable.
\\
\\
Firstly, it should be valuable to check the random distribution of Taylor coefficients. We consider some lowest expansions of Taylor series as examples.
\begin{align}
&{{A}_{h}}(x,y)={{a}_{00}}+{{a}_{10}}(x-{{x}_{0}})+{{a}_{01}}(y-{{y}_{0}})+{{a}_{02}}{{(x-{{x}_{0}})}^{2}}+\nonumber\\
&{{a}_{20}}{{(y-{{y}_{0}})}^{2}}+{{a}_{11}}(x-{{x}_{0}})(y-{{y}_{0}})+\ldots
\end{align}
The following Figure \ref{distri} shows the probability distribution of some Taylor coefficients as a sample. One can see that they are both roughly symmetrically distributed around zero, while the sine activation has a larger random range, means that sine activation is more random in this range\footnote{If we use non-monotonic activation function in the neural network to do machine learning, the output is relatively chaotic and might be relatively hard to converge, especially when we use the back propagation algorithms. However, practically it is more beneficial for our random function generation.}.
\begin{figure}
\centering
\includegraphics[width=0.8\textwidth]{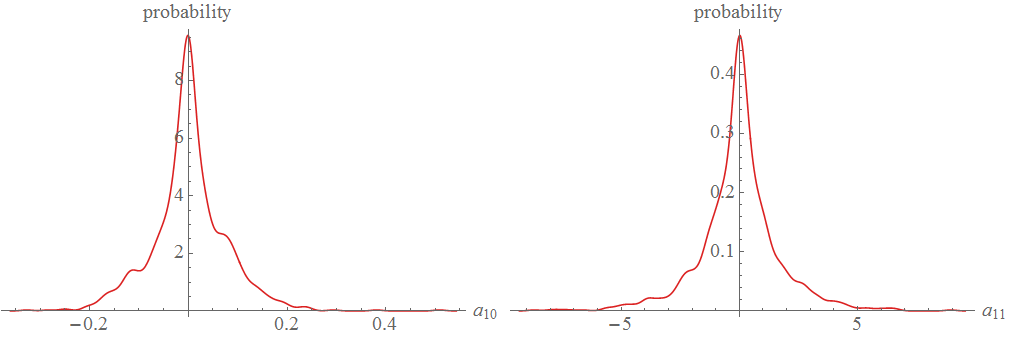}\\
\includegraphics[width=0.8\textwidth]{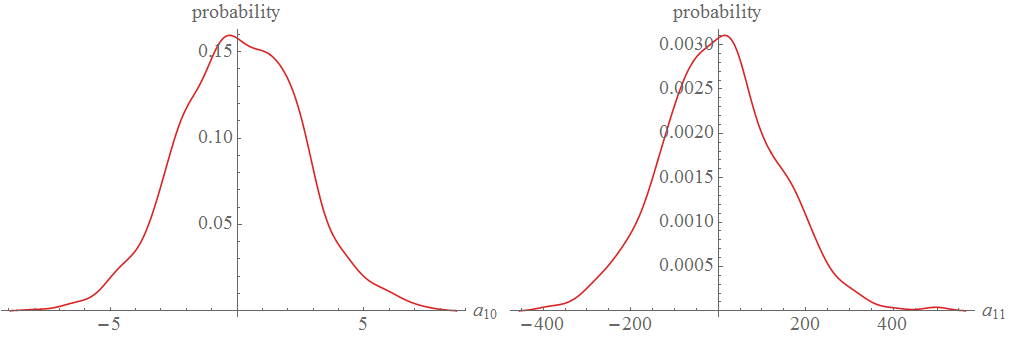}
\caption{The probability distribution of the Taylor series coefficients $a_{10}$ and $a_{11}$ for neural network in two dimension. We consider the sigmoid activation (upper) and the sine activation (lower) respectively. The data is constructed via $x_0=y_0=1$, $\alpha=100$, $\beta=100$, $w=1/h$ and $h=100$, and we realize the network for 1000 times.} \label{distri}
\end{figure}
\\
\\
Secondly, in order to do some simple tests to probe some hidden correlations between different Taylor coefficients, a nice type of statistical quantities is \emph{covariance}
\begin{align}
\text{Cov}(X,Y)=\frac{1}{\mathcal{N}}\sum\limits_{i=1}^{\mathcal{N}}{({{X}_{i}}-\bar{X})({{Y}_{i}}-\bar{Y})}
\end{align}
and the \emph{correlation}
\begin{align}
\eta (X,Y)=\frac{\text{Cov}(X,Y)}{\sqrt{\text{Cov}(X,X)\text{Cov}(Y,Y)}}
\end{align}
for random variable $X$ and $Y$ with number of samples $\mathcal{N}$. Those quantities are standard statistical justification of the correlation (or more precisely, linear correlation) of some random variables. When the absolute value of $\eta$ is near 1, then variables are strongly correlated. When the absolute value of $\eta$ is approaching zero, then variables have weak or no (linear) correlation.
\\
\\
In the following Figure \ref{corr}, we give the plots of the absolute values of the correlations between different Taylor coefficients. Based on this analysis, we could conclude the following results,
\begin{itemize}
\item We do find that several Taylor coefficients are strongly correlated ($|\eta|\ge 0.5$). However, there are also several Taylor coefficients that are weakly correlated ($|\eta|<0.1$). The categories of coefficients justifying correlation, are activation function dependent. Namely, it is not simply due to the structure of neural networks.
\item One can turn up and down several parameters of the network. However, most parameters justifying random ranges in the network are effectively rescaling the input and output variables, except the number of hidden neurons, $h$, parametrizing the complexity of the network. Thus, in the plots we test the dependence between the correlation and the variable $h$. However, the dependence is not clear at least in our testing range. Maybe it is because the $h$ in our setup is too small. At larger scale of $h$, the correlation might change dramatically. However, by our computational resource it is hard to make a concrete statement.
\item There are some other issues about this numerical tests. For instance, the correlation might also depend on the location when we do Taylor expansion, the shape and structure of neural networks, and the input measure of random variables. And one can also seek for higher order correlations instead of linear. We leave those issues to future research.
\end{itemize}
\begin{figure}
\centering
\includegraphics[width=0.8\textwidth]{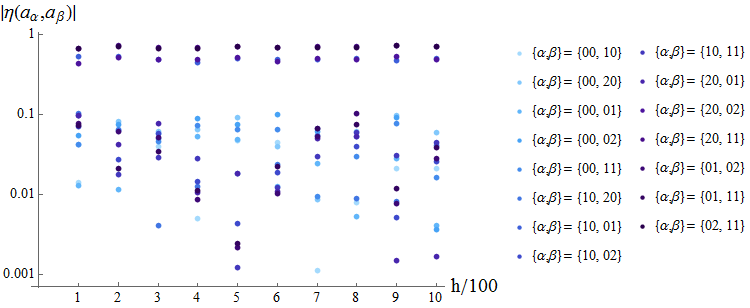}\\
\includegraphics[width=0.8\textwidth]{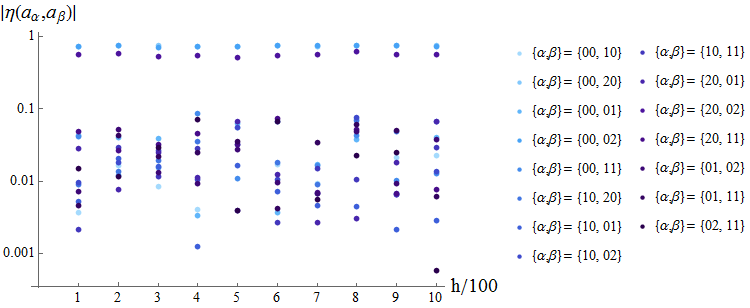}
\caption{The absolute value of (linear) correlations between Taylor series coefficients.  We consider the sigmoid activation (upper) and the sine activation (lower) respectively. The data is constructed via $x_0=y_0=1$, $\alpha=100$, $\beta=100$, $w=1/h$ and $h$ changes from 100 to 1000, and we realize the network for 1000 times.} \label{corr}
\end{figure}

\subsection{Cosmological application}
As a simple application, we will test a concrete inflation model. Considering multiple fields living in a high dimensional landscape, the Friedman equation is written as
\begin{align}
\ddot{\varphi_i}+3H\dot{\varphi_i}+\partial_iV(\varphi_1,\varphi_2\ldots,\varphi_n)=0
\end{align}
where $i=1,2\ldots,n$ for $n$ fields. In inflation the Hubble parameter would be approximated as
\begin{align}
H=\sqrt{\frac{V(\varphi )}{3}}
\end{align}
in the slow roll limit. This equation the classical dynamics of the background fields. To include quantum fluctuation, Starobinsky method \cite{Starobinsky:1986fx} is typically used for numerical simulations of an inflationary trajectory. Integrating out the sub-horizon modes, quantum fluctuations could be effectively serving as a source term in the equation of motion, namely
\begin{align}
\ddot{\varphi_i}+3H\dot{\varphi_i}+\partial_iV = \frac{3}{2\pi}H^\frac{5}{2}\eta_i(t)
\end{align}
where the $\eta_i$ term is the random source to denote the quantum corrections and follows independent Gaussian distribution. We construct the normalization condition to constrain the $\eta_i$ term,
\begin{align}
\langle \eta_i(t)\eta_{j}(t')\rangle=\delta(t-t')\delta_{ij}
\end{align}
such that during a Hubble time and averaged on a Hubble volume, the quantum fluctuation on each field direction is normalized as $H/(2\pi)$. In numerical calculation, the discrete time interval $\Delta t=t_k-t_{k-1}$ has to be used. While replacing differential equations into difference equations, the Dirac delta function shall be replaced by
\begin{align}
\delta(t-t')\to\frac{\delta_{kk'}}{\Delta t}
\end{align}
So the difference strategy is
\begin{align}
\frac{{{\varphi }_{i}}(k+1)+{{\varphi }_{i}}(k-1)-2{{\varphi }_{i}}(k)}{2\Delta {{t}^{2}}}+3\sqrt{\frac{V({{\varphi }_{j}}(k))}{3}}\frac{{{\varphi }_{i}}(k+1)-{{\varphi }_{i}}(k)}{\Delta t}+{{\partial }_{i}}V({{\varphi }_{j}}(k))=\frac{{3{H}^{5/2}}}{2\pi }{{\eta }_{i}}
\end{align}
where $\eta_i(k)\sim N(0,1/\sqrt{\Delta t})$, and Gaussian distributions are independent for different $i$.
\\
\\
In our simple model, we will assume that inflation is mainly only by one field $\varphi_1$, and the potential looks like
\begin{align}
V={{V}_{0}}({{\varphi }_{1}})+{{V}_{\mathrm{rand}}}(\varphi_1,{{\varphi }_{2}},\ldots ,{{\varphi }_{n}})
\end{align}
where $V_0$ would be chosen to be quadratic (large field inflation)
\begin{align}
{{V}_{0}}({{\varphi }_{1}})=\frac{1}{2}{{m}^{2}}\varphi _{1}^{2}
\end{align}
The random part $V_{\mathrm{rand}}$, is generated by neural networks with some random amplitudes $\mathcal{A}$. As a simple example, we choose the parameters as $n=20$ (for 20-dimensional field space), $m=10^{-6}$ (inflaton mass normalized by Planck mass), $\varphi_{\max}=30$ (the initial value and bounded value of fields), $\Delta t=10^5$ (discrete time scale). The initial value of fields are: $\varphi_1=\varphi_{\max}$, $\varphi_{i\ne1}=\varphi_{\max}/2$. During the evolution of the trajectory, if field one field drops out of the range $[0, \varphi_{\max}]$, we think it as the end of inflation and breaks the recursion.
\begin{figure}[htbp]
  \centering
  \includegraphics[width=0.4\textwidth]{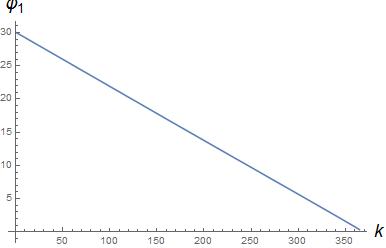}
  \includegraphics[width=0.4\textwidth]{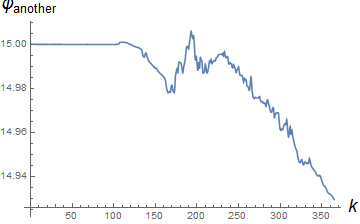}
  \caption{\label{fig:log} The evolution of trajectories for dimension 20. In this case we choose the sigmoid activation function and we set $\alpha=\beta=50$, $\omega=1$, $\mathcal{A}=10^{-14}$ and $h=100$. And also we set $V_\text{rand}=A_h(\varphi_1-\varphi_\text{max}/2,(\varphi_{i\ne 1}-\varphi_\text{max}/2)\times 10)$. We show the trajectories $\varphi_1(k)$ (left) as a function of time step $k$ for inflaton, and  $\varphi_\text{another}(k)$ (left) for another light field.}
\end{figure}
\\
\begin{figure}[htbp]
  \centering
  \includegraphics[width=0.4\textwidth]{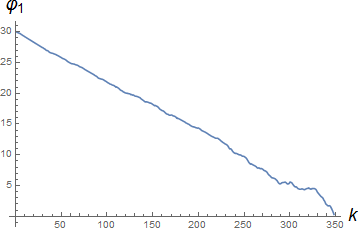}
  \includegraphics[width=0.4\textwidth]{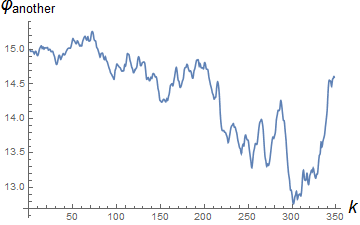}
  \caption{\label{fig:sine} The evolution of trajectories for dimension 20. In this case we choose the sine activation function and we set $\alpha=6000$, $\beta=100$, $\omega=1$, $\mathcal{A}=10^{-15}$ and $h=100$. And also we set $V_\text{rand}=A_h(\varphi_1,\varphi_{i\ne 1})$. We show the trajectories $\varphi_1(k)$ (left) as a function of time step $k$ for inflaton, and  $\varphi_\text{another}(k)$ (left) for another light field.}
\end{figure}
\\
We can see some expected features in Figure \ref{fig:log} and Figure \ref{fig:sine}. The behavior of $\varphi_1$ is roughly linear, namely the characteristic slow-roll inflation with the quadratic potential, which supports that in this model, the quadratic potential dominates the whole process of inflation. We try different forms of activation functions, and the results are different. The sine activation function makes more random potential (as discussed previously), thus the vibration is relatively stronger for light fields other than the inflaton. These simulation could be used for further study for dynamics of inflation and compare to observations.
\section{Conclusion and discussion}\label{conc}
In this paper we show that artificial neural network is useful to construct a cosmic landscape for computer simulation. After a simple discussion on the computational complexity of random functions, we introduce artificial neural network, the foundation of machine learning and make the training parameter to be random in order to construct random functions. We show that neural network construction of a random landscape is fast up to the construction of hidden layers and number of hidden neurons. Finally, we present an example to use such a random landscape to simulate the dynamics of inflationary trajectories, where the dimension of field space is hard to achieve for ordinary methods.
\\
\\
We are trying to show the possible computational power for neural network technologies in the application of high-dimensional cosmological landscape. Some further researches could be done in the future. Firstly, in this paper we choose a multilayer perceptron with one hidden-layer for simplicity, but in the future, to increase the complexity of random potential one should consider neural networks with more complicated structures, like more hidden layers and recurrent neural networks. Furthermore, more cosmological models and their properties could be addressed and simulated with neural networks when considering landscape potentials. We leave these interesting possibilities to future works.

\section*{Acknowledgments}
JL thanks the instructor Zejun Ding for his help and encouragement when some of ideas are presented in a final presentation of a course project of his course Computational Physics A in 2014 fall, when the author was a college student in the University of Science and Technology of China. We thank Sean Carroll, Ashmeet Singh and Yi Wang for their communications, Hao Xu for his valuable supports on computer science and programming, and Jiahui Liu for her discussions and initial collaboration on this project. JL is supported in part by the Walter Burke Institute for Theoretical Physics in California Institute of Technology, by U.S. Department of Energy, Office of Science, Office of High Energy Physics, under Award Number DE-SC0011632.

\end{document}